\documentclass[manuscript]{acmart}
\AtBeginDocument{%
  \providecommand\BibTeX{{%
    \normalfont B\kern-0.5em{\scshape i\kern-0.25em b}\kern-0.8em\TeX}}}

\usepackage{multirow}
\usepackage{amsmath}
\usepackage{mathtools}
\usepackage{graphicx}
\usepackage{listings}
\usepackage{xcolor}
\usepackage{verbatim}
\usepackage{minted}
\usepackage{tcolorbox}
\tcbuselibrary{breakable}
\usepackage[utf8]{inputenc}
\usepackage{xcolor}
\usepackage{hyperref}
\usepackage{enumitem}
\usepackage{tikz}
\usepackage{graphicx}
\usepackage{tabularx}
\usepackage{booktabs}

\begin{document}

\title[Beyond Semantic Similarity: Open Challenges for Embedding-Based Creative Process Analysis]{Beyond Semantic Similarity: Open Challenges for Embedding-Based Creative Process Analysis Across AI Design Tools}

\author{Seung Won Lee}
\authornote{Both authors contributed equally to this work.}
\email{lswgood0901@gmail.com}
\affiliation{%
  \institution{Design Informatics Lab, Hanyang University}
  \city{Seoul}
  \postcode{04763}
  \country{Republic of Korea}
}
\affiliation{%
  \institution{Human-Centered AI Design Institute, Hanyang University}
  \city{Seoul}
  \country{Republic of Korea}
}

\author{Semin Jin}
\authornotemark[1]
\email{tpals97@gmail.com}
\affiliation{%
  \institution{Design Informatics Lab, Hanyang University}
  \city{Seoul}
  \postcode{04763}
  \country{Republic of Korea}
}
\affiliation{%
  \institution{Human-Centered AI Design Institute, Hanyang University}
  \city{Seoul}
  \country{Republic of Korea}
}

\author{Kyung Hoon Hyun}
\authornote{Corresponding author.}
\email{hoonhello@gmail.com}
\affiliation{%
  \institution{Design Informatics Lab, Hanyang University}
  \city{Seoul}
  \postcode{04763}
  \country{Republic of Korea}
}
\affiliation{%
  \institution{Human-Centered AI Design Institute, Hanyang University}
  \city{Seoul}
  \country{Republic of Korea}
}

\renewcommand{\shortauthors}{Seung Won Lee, Semin Jin, and Kyung Hoon Hyun}

\begin{abstract}
AI-based creativity support tools (CSTs) are evaluated through domain-specific metrics, limiting cross-domain comparison of creative processes. Embedding-based protocol analysis offers a potential domain-agnostic analytical layer. However, we argue that fixed embedding similarity can misrepresent creative dynamics: it may not detect creative pivots that occur within superficially similar language, treating shifts in the problem being addressed as continued elaboration. We identify three open challenges stemming from this gap: aligning similarity measures with creative significance, segmenting and representing multimodal design traces, and evaluating agentic systems where embedding-based metrics enter the generation loop and shape agent behavior. We propose context-aware interventions using large language models as a direction for making trace analysis sensitive to session-specific creative dynamics.

\end{abstract}

\keywords{creative activity traces, creativity support tools, design process analysis, linkography, AI-mediated creativity}

\maketitle

\section{Introduction}

AI-based creativity support tools (CSTs) are evaluated through domain-specific metrics, making findings difficult to compare across systems. This paper asks how such cross-domain comparison might be achieved, and what obstacles remain.

We argue that cross-domain comparison of creative processes requires shifting from content-level evaluation---judging the quality or characteristics of final design outputs---to process-level structural analysis that examines how creative exploration unfolds over time: how designers branch into new directions, revisit earlier ideas, and develop concepts through iteration. Embedding-based trace analysis offers a promising approach. It takes the sequence of discrete design actions that a designer performs during a design process and projects them into a shared representational space using neural embedding models. From these representations, one can construct a fuzzy linkograph~\cite{smith2025fuzzy}, a directed graph that maps semantic relationships between design moves, and derive metrics such as entropy and link density that characterize the process independently of domain-specific content. This reasoning finds support in recent work on recommendation evaluation: Ishii et al.~\cite{ishii2026bigpu} showed that conventional similarity-based metrics become unreliable across systems because they conflate item familiarity with meaningful exploration, and that an information-theoretic formulation enables more robust cross-system comparison. Creative process evaluation faces an analogous challenge: fixed embedding models measure surface-level semantic overlap between moves, so they cannot distinguish genuine conceptual continuity from creative pivots that happen to share similar vocabulary—the kind of subtle but significant shifts that often drive a design process forward.

However, realizing this vision is far from straightforward. This paper identifies three open challenges at the intersection of representation and evaluation: (1)~whether the similarity structures captured by current embedding models align with what matters creatively, (2)~how such analysis can accommodate multimodal design traces, and (3)~how to evaluate creative processes in increasingly agentic AI systems where both humans and AI contribute to the creative trajectory.

\section{From Siloed Evaluations to Process-Level Analysis}
Current CST evaluations lack a principled basis for comparing process structures across tools. Subjective instruments such as the Creativity Support Index (CSI)~\cite{cherry2014quantifying} and NASA-TLX~\cite{hart1988development}—employed, for instance, by GenQuery~\cite{son2024genquery} and InkSpire~\cite{lin2025inkspire} to assess creative support in visual search and product design, respectively—capture perceived experience. \textit{Manual process analysis}, as in Lee et al.'s~\cite{lee2024impact} linkographic comparison of sketch- and prompt-guided 3D modeling, can reveal modality-specific process dynamics but requires significant human effort and resists standardization. \textit{Domain-specific quantitative metrics}, such as FontCraft's~\cite{tatsukawa2025fontcraft} convergence measures for optimization-driven font design, are tightly coupled to particular workflows. None of these approaches addresses this gap.

Fuzzy-linkography~\cite{smith2025fuzzy} offers a potential path forward. It automates linkograph construction~\cite{goldschmidt2014linkography} by computing semantic similarity between sequential design moves using embedding models, deriving established metrics such as entropy, link density, and critical moves~\cite{kan2007quantitative}. The key property is that it operates on traces of design activity rather than on domain-specific outputs: any CST producing recordable design moves---prompts, sketches, selections, or parameter adjustments---could in principle be analyzed through this lens. However, whether such analysis yields valid cross-tool comparisons depends on several assumptions that remain untested.

\section{Open Challenges and Research Agenda}

Realizing this potential requires addressing several open questions that we propose as a research agenda for the community.

\subsection{Semantic Similarity vs. Creative Significance}

General-purpose embeddings measure semantic overlap between utterances, but in design processes, creative significance often resides in shifts that occur within superficially similar language. Consider a designer working on a micro-apartment project who starts with \textit{"stackable chair modules for compact storage"}—solving the problem of limited floor space when furniture is not in use. The designer then proposes \textit{"stackable wall modules for reconfigurable room layouts,"} realizing that the same principle could let residents transform a studio into separate sleeping and working areas on demand. An embedding model would score these moves as highly similar—both share \textit{"stackable"} and \textit{"modules"}—but the second move is a creative pivot: the designer has reframed a furniture storage strategy as a solution to an entirely different problem, spatial adaptability in constrained housing. Under embedding-based trace analysis, such moves would be linked as continuous elaboration of a single idea, when in fact they mark a shift in the problem being addressed. The resulting link structure would understate the exploratory breadth of the session, and derived metrics such as entropy would reflect apparent conceptual continuity rather than the actual dynamics of ideation.

Large language models offer several possible intervention points: preprocessing raw interaction logs to segment design moves around meaningful conceptual shifts; assisting link formation by judging whether two moves share creative relevance given the session's task framing and preceding trajectory; or reinterpreting multimodal traces by extracting design intent from sketches or images that fixed embeddings would reduce to visual similarity alone. Whether such context-aware interventions produce link structures that better align with expert judgment remains an empirically testable question, and we see their investigation as a concrete next step toward closing the gap between system-computed and creativity-relevant measures.

\subsection{Multimodal Trace Integration}

Most CSTs involve multimodal interaction, yet current automated trace analysis focuses primarily on text. While multimodal embedding models such as CLIP~\cite{radford2021learning} could extend fuzzy-linkography to visual modalities, computing similarity between images does not straightforwardly translate to measuring creative process dynamics. Two images might be visually similar yet represent different design strategies, or visually dissimilar yet represent a coherent creative evolution---a rough early sketch and a refined final rendering may share little visual similarity while representing continuous development of the same concept.

Moreover, the meaningful unit of analysis---what constitutes a ``design move''---becomes more ambiguous in multimodal contexts. In text-based interaction, a prompt-response pair provides a natural segmentation boundary. But in InkSpire's~\cite{lin2025inkspire} sketch-and-generate cycles, a single design move might span a sequence of pen strokes, a generation request, and a selective incorporation of AI suggestions. In FontCraft's~\cite{tatsukawa2025fontcraft} preference-based optimization, each feedback iteration involves an implicit comparison across multiple candidates---is this one move or several? Without principled segmentation methods, the resulting linkographic structures may reflect arbitrary analysis choices rather than meaningful process dynamics. Developing domain-informed but generalizable segmentation heuristics is a prerequisite for extending embedding-based analysis beyond text.

\subsection{Evaluating Creative Processes in Agentic AI Systems}
As CSTs evolve toward agentic architectures—where AI agents autonomously generate alternatives and steer creative direction—embedding-based metrics take on a different role. Agents must evaluate their own design trajectories in real time to decide what to propose next, when to shift direction, and when to continue elaborating. If these evaluations rely on embedding similarity, the limitations described above no longer sit at the analysis stage—they enter the generation loop itself, shaping what the agent produces.

This compounds the evaluation problem. An agent's generation policy—its diversity settings, sampling strategy, and steering heuristics—will leave signatures in any process trace. An agent configured to maximize output variety may produce traces with high link entropy that reflect its policy rather than genuine creative development. Without methods to disentangle such artifacts from the designer's creative dynamics, process metrics cannot reliably assess whether an agentic CST is a productive collaborator or one that merely produces varied output. Developing such methods might require comparative studies across different levels of agent autonomy on matched design tasks—an empirical agenda the community has yet to undertake.

\section{Conclusion}

We have argued that embedding-based trace analysis offers a promising direction for cross-domain comparison of creative processes in AI-based CSTs, but that its utility depends on closing the gap between what fixed embedding similarity captures and what constitutes creative significance for human designers. This gap manifests in three dimensions: in the failure to detect creative pivots that occur within superficially similar language, in the added ambiguity of segmenting and representing multimodal design activity, and in agentic systems where embedding-based evaluation enters the generation loop and shapes the agent's creative behavior itself. We propose that large language models, leveraged as context-aware intermediaries in the analytical pipeline, offer a viable path toward making trace analysis sensitive to session-specific creative dynamics, and that comparative studies across different levels of AI agency could provide the empirical grounding this agenda requires.

\section*{Acknowledgments}
This work was supported by the Industrial Technology Innovation Program(RS-2025-02317326, Development of AI-Driven Design Generation Technology Based on Designer Intent) funded by the Ministry of Trade, Industry \& Energy (MOTIE, Korea).

\bibliographystyle{ACM-Reference-Format}
\bibliography{references}

\end{document}